\documentclass[aps, pra, showpacs, reprint]{revtex4-1}

\begin{document}

\title{Quantum discord for two-qubit X states: Analytical formula with very small worst-case error}

\author{Yichen Huang}

\email{yichenhuang@berkeley.edu}

\affiliation{Department of Physics, University of California, Berkeley, Berkeley, California 94720, USA}

\date{\today}

\begin{abstract}

Quantum discord is a measure of quantum correlation beyond entanglement. Computing quantum discord for simple quantum states is a basic problem. An analytical formula of quantum discord for two-qubit X states is first claimed in [Ali, Rau, and Alber, Phys. Rev. A 81, 042105 (2010)], but later found to be not always correct. I observe numerically that the formula is valid with worst-case absolute error 0.0021. For symmetric two-qubit X states, I give a counterexample to the analytical formula derived in [F. F. Fanchini et al., Phys. Rev. A 81, 052107 (2010)], but observe that the formula is valid with worst-case absolute error 0.0006. The formula has been used in many research papers. The results in all these works are approximately correct, even if they may not be exactly correct.

\end{abstract}


\maketitle

Quite a few fundamental concepts in quantum mechanics do not have classical analogs: uncertainty relations \cite{Rob29, Bec75, BM75, Hua11, *Hua12}, quantum nonlocality \cite{EPR35, CHSH69, PV07, HHHH09}, etc. Quantum entanglement is defined based on the notion of local operations and classical communication (LOCC): a bipartite quantum state is separable (not entangled) if it can be created by LOCC \cite{PV07, HHHH09}. The set of separable states is convex and has nonzero measure (volume), and a lot of effort is devoted to entanglement detection \cite{HHHH09, Per96, DGCZ00, Sim00, GT09, Hua10, *Hua10E, *Hua13}. However, it is argued that nontrivial quantum correlation also exists in certain separable states. A number of measures have been reported to quantify quantum correlation beyond entanglement \cite{MBC+12}. Quantum discord, proposed explicitly in \cite{OZ01} and implicitly in \cite{HV01}, is the most popular such measure and a hot research topic in the past a few years. The set of classical (zero discord) states is nowhere dense and has zero measure (volume) \cite{FAC+10}. Unfortunately, computing quantum discord seems extremely difficult as its definition (\ref{definition}) requires the optimization over all measurements. Few analytical results are known even for two-qubit states, and the computational cost of any numerical approach is expected to grow exponentially with the dimension of the Hilbert space.

Let us focus on two-qubit X states, which we frequently encounter in condensed matter systems, quantum dynamics, etc. \cite{FWB+10, Dil08, Sar09, WTRR10, CRC10}. For instance, the two-site reduced density matrix of the symmetry unbroken ground state of a lattice Hamiltonian with $Z_2$ symmetry is of the X structure (\ref{X}). An analytical formula of quantum discord for Bell-diagonal states (a subset of two-qubit X states) is known \cite{Luo08}. For general two-qubit X states, the first attempt is made in \cite{ARA10, *ARA10E}, and the analytical formula (\ref{discord}) is claimed. However, (\ref{discord}) is not always correct: a counterexample is given in \cite{LMXW11} (see also \cite{CZY+11}). The reason is that not all extrema are identified in \cite{ARA10, *ARA10E}, and not all constraints are taken into consideration \cite{LMXW11}. Hence, the analytical formula of quantum discord for general two-qubit X states is still unknown. Two regions in which (\ref{discord}) is valid are identified in \cite{CZY+11}. Moreover, there are statistical evidences that (\ref{discord}) and related formulae are pretty good approximations for most states \cite{LMXW11, GGZ11, VR12, QQJ12}, although these evidences do not rule out the possibility that an unlucky guy obtains qualitatively incorrect results for some states by using these formulae. Symmetric two-qubit X states are of special interest in condensed matter systems, quantum dynamics, etc. \cite{FWB+10, Dil08, Sar09, WTRR10, CRC10}. For instance, the two-site reduced density matrix of the symmetry unbroken ground state of a translationally invariant lattice Hamiltonian is symmetric. An analytical formula, which happens to be equivalent to (\ref{discord}), for this subset of two-qubit X states is derived independently in \cite{FWB+10}.

Computing quantum discord numerically for two-qubit X states is straightforward. Surprisingly, I observe that even if (\ref{discord}) is not always correct exactly, it is always correct approximately with very small worst-case absolute error (\ref{result1}). Technically, I search over the entire space of two-qubit X states with steps small enough to ensure numerical precision. For symmetric two-qubit X states, (in contrast to \cite{FWB+10}) (\ref{discord}) is still not always correct [(\ref{counterexample2}) gives a counterexample], again because not all extrema are identified in \cite{FWB+10}. In this case, as expected the worst-case absolute error is smaller (\ref{result2}). Equation (\ref{discord}) has been used in many (about 80) research papers: e.g., \cite{FWB+10, ARA10, CLSM10, AD10, GGZ11a, *GGZ11E, LWF11, WRR11} (I do not list them all here). The results in all these works are approximately correct, even if they may not be exactly correct.


\section{technical perspective}

Mutual information in classical information theory has two inequivalent quantum analogs. Quantum mutual information $I(\rho_{AB})=S(\rho_A)+S(\rho_B)-S(\rho_{AB})$ quantifies the total correlation of the bipartite quantum state $\rho_{AB}$, where $S(\rho_A)=-\text{tr}\rho_A\ln\rho_A$ is the von Neumann entropy of the reduced density matrix $\rho_A=\text{tr}_B\rho_{AB}$. Let $\{\Pi_i\}$ be a measurement on the subsystem $B$. Then, $p_i=\text{tr}(\Pi_i\rho_{AB})$ is the probability of the $i$th measurement outcome, and $\rho_A^i=\text{tr}_B(\Pi_i\rho_{AB})/p_i$ and $\rho'_{AB}=\sum_ip_i\rho_A^i\otimes\Pi_i$ are post-measurement states. Classical correlation is defined as $J_B(\rho_{AB})=\max_{\{\Pi_i\}}J_{\{\Pi_i\}}(\rho_{AB})$, where $J_{\{\Pi_i\}}(\rho_{AB})=S(\rho_A)-\sum_ip_iS(\rho_A^i)$ \cite{HV01}. The maximization is taken either over all von Neumann measurements or over all generalized measurements described by positive-operator valued measures (POVM). For simplicity, we restrict ourselves to von Neumann measurements in this work. Quantum discord, a measure of quantum correlation beyond entanglement, is the difference between total correlation and classical correlation \cite{OZ01}:
\begin{eqnarray}
D_B(\rho_{AB})&=&\min_{\{\Pi_i\}}D_{\{\Pi_i\}}(\rho_{AB})=I(\rho_{AB})-J_B(\rho_{AB})\nonumber\\
&=&S(\rho_B)-S(\rho_{AB})+\min_{\{\Pi_i\}}\sum_ip_iS(\rho_A^i)\nonumber\\
&=&\min_{\{\Pi_i\}}S_B(\rho_{AB}')-S_B(\rho_{AB}),
\label{definition}
\end{eqnarray}
where $S_B(\rho_{AB})=S(\rho_{AB})-S(\rho_B)$ is the quantum conditional entropy, and $D_{\{\Pi_i\}}(\rho_{AB})=I(\rho_{AB})-J_{\{\Pi_i\}}(\rho_{AB})$. 
Quantum discord is invariant under local unitary transformations.

Labeling the basis vectors $|1\rangle=|00\rangle,|2\rangle=|01\rangle,|3\rangle=|10\rangle,|4\rangle=|11\rangle$, the density matrix of a two-qubit X state
\begin{equation}
\rho_{AB}=\left(\begin{array}{cccc}a&0&0&\alpha\\0&b&\beta&0\\0&\bar\beta&c&0\\\bar\alpha&0&0&d\end{array}\right)
\label{X}
\end{equation}
has nonzero elements only on the diagonal and the antidiagonal, where $a,b,c,d\ge0$ satisfy $a+b+c+d=1$, and the positive semidefiniteness of $\rho_{AB}$ requires $|\alpha|^2\le ad,|\beta|^2\le bc$. The antidiagonal elements $\alpha,\beta$ are generally complex numbers, but can be made real and nonnegative by the local unitary transformation $e^{-i\theta_1\sigma_z}\otimes e^{-i\theta_2\sigma_z}$ with suitable $\theta_1,\theta_2$, where $\sigma$ is the Pauli matrix; assume without loss of generality $\alpha,\beta\ge0$. Hereafter I follow and generalize the approach of \cite{CRC10}. Parametrizing a von Neumann measurement $\{\Pi_i=|i'\rangle\langle i'|\}$ by two angles $\theta,\phi$: $|0'\rangle=\cos(\theta/2)|0\rangle+e^{i\phi}\sin(\theta/2)|1\rangle$ and $|1'\rangle=\sin(\theta/2)|0\rangle-e^{i\phi}\cos(\theta/2)|1\rangle$, (\ref{definition}) is reduced to a minimization over two variables. The eigenvalues of the post-measurement state $\rho_{AB}'$ are
\begin{eqnarray}
\lambda_{1,2}&=&\{1+(a-b+c-d)\cos\theta\nonumber\\
&&\pm[(a+b-c-d+(a-b-c+d)\cos\theta)^2\nonumber\\
&&+4(\alpha^2+\beta^2+2\alpha\beta\cos2\phi)\sin^2\theta]^{1/2}\}/4,\nonumber\\
\lambda_{3,4}&=&\{1-(a-b+c-d)\cos\theta\nonumber\\
&&\pm[(a+b-c-d-(a-b-c+d)\cos\theta)^2\nonumber\\
&&+4(\alpha^2+\beta^2+2\alpha\beta\cos2\phi)\sin^2\theta]^{1/2}\}/4;
\end{eqnarray}
the eigenvalues of $\rho_B'$ are $\Lambda_{1,2}=(1\pm(a-b+c-d)\cos\theta)/2$. We would like to minimize the quantum conditional entropy
\begin{equation}
S_B(\rho_{AB}')=\Lambda_1\ln\Lambda_1+\Lambda_2\ln\Lambda_2-\sum_{i=1}^4\lambda_i\ln\lambda_i.
\label{max}
\end{equation}
Thanks to the concavity of the Shannon entropy, the minimization over $\phi$ can be worked out exactly: $\cos2\phi=1$. Indeed, one can verify $\partial S_B(\rho_{AB}')/\partial\cos(2\phi)\le0$. Then, a single-variable minimization suffices. There are at least two extrema: $\theta=0,\pi/2$, and the measurements are $\sigma_z,\sigma_x$, respectively. It is tempting (but not always correct) to write down the analytical formula:
\begin{equation}
D_B(\rho_{AB})=\min\{D_{\sigma_x}(\rho_{AB}),D_{\sigma_z}(\rho_{AB})\}.
\label{discord}
\end{equation}
As is rephrased by \cite{CZY+11}, (\ref{discord}) is equivalent to the main result of \cite{ARA10}, obtained in a different approach. In the case that $a=d,b=c$ the algebra is greatly simplified; the validity of (\ref{discord}) can be verified explicitly; the main result of \cite{Luo08} is reproduced.

Setting $\phi=0$, the single-variable expression (\ref{max}) we would like to minimize (over $\theta$) is lengthy and complicated. This is strong evidence that for general two-qubit X states quantum discord cannot be evaluated analytically, even if it is straightforward to compute numerically. Surprisingly, only a very small absolute error occurs when there are additional extrema besides $\theta=0,\pi/2$:
\begin{equation}
D_B(\rho_{AB})>\min\{D_{\sigma_x}(\rho_{AB}),D_{\sigma_z}(\rho_{AB})\}-0.0021.
\label{result1}
\end{equation}
Technically, for our purpose the density matrix $\rho_{AB}$ can be parametrized by four free parameters: $a,b,c,d$ with the constraint $a+b+c+d=1$ and $\alpha+\beta$ as $\alpha+\beta$ appears as a combination in (\ref{max}); flipping the first qubit and/or the second qubit if necessary, assume without loss of generality $a+b\le c+d$ and $a\ge b$. This reduced space of two-qubit X states is searched over with different steps in different regions for efficient use of computational resources; the steps are kept very small, e.g., $10^{-6}$ in the vicinity of (\ref{counterexample1}), in the region the absolute error of (\ref{discord}) is large to ensure numerical precision. The state with the largest absolute error 0.002047 (and $\theta=0.607573$) I find is
\begin{equation}
\rho_{AB}=\left(\begin{array}{cccc}0.027180&0&0&0.141651\\0&0.000224&0&0\\0&0&0.027327&0\\0.141651&0&0&0.945269\end{array}\right).
\label{counterexample1}
\end{equation}

A two-qubit X state $\rho_{AB}$ is symmetric if $\rho_A=\rho_B$ or $b=c$ in (\ref{X}). For symmetric two-qubit X states, I observe
\begin{equation}
D_B(\rho_{AB})>\min\{D_{\sigma_x}(\rho_{AB}),D_{\sigma_z}(\rho_{AB})\}-0.0006
\label{result2}
\end{equation}
by similar numerical analysis with one fewer free parameter. The analytical formula derived independently in \cite{FWB+10}, which happens to be equivalent to (\ref{discord}), is also not always correct. The state with the largest absolute error 0.000573 (and $\theta=0.477918$) I find is
\begin{equation}
\rho_{AB}=\left(\begin{array}{cccc}0.021726&0&0&0.128057\\0&0.010288&0&0\\0&0&0.010288&0\\0.128057&0&0&0.957698\end{array}\right).
\label{counterexample2}
\end{equation}

\begin{acknowledgements}

The author would like to thank Joel. E. Moore for useful suggestions, and Felipe F. Fanchini and Nicolas Quesada for discussions. This work was supported by the United States Army Research Office via the Defense Advanced Research Projects Agency--Optical Lattice Emulator program.

\end{acknowledgements}

\bibliography{discord}

\end{document}